\documentclass[11pt,twoside]{article}

\usepackage{asp2006}
\usepackage{graphicx}
\usepackage{lscape}

\begin{document}
\title{Fermi/LAT discovery of gamma-ray emission from a relativistic jet in the narrow-line Seyfert 1 quasar PMN~J0948+0022}   

\author{L. Foschini\altaffilmark{1} on behalf of the \emph{Fermi}/LAT Collaboration\\and\\G. Ghisellini\altaffilmark{1}, L. Maraschi\altaffilmark{1}, F. Tavecchio\altaffilmark{1}, E. Angelakis\altaffilmark{2}}

\affil{$^{\rm 1}$INAF - OA Brera, Via E. Bianchi 46 - 23807 Merate (LC) - Italy}
\affil{$^{\rm 2}$MPI Radioastronomie, Auf dem H\"ugel 69 - 53121 Bonn - Germany}    

\begin{abstract} 
We report the discovery by the Large Area Telescope (LAT) onboard the \emph{Fermi Gamma-ray Space Telescope} of high-energy gamma-ray emission from the peculiar quasar PMN~J0948+0022 ($z=0.585\pm 0.001$). Contrary to the expectations, the optical spectrum of this quasar shows only narrow lines [FWHM($H\beta$) $\sim 1500$~km/s] and the typical characteristics of narrow-line Seyfert 1 type galaxies. However, the strong radio emission and the flat spectrum suggest the presence of a relativistic jet, which can now be confirmed by the detection of MeV-GeV photons. PMN~J0948+0022 is therefore the first radio-loud narrow-line Seyfert 1 quasar to be detected at gamma-rays and the third type of $\gamma-$ray emitting AGN, after blazars and radiogalaxies.
\end{abstract}

\section{Introduction}   
It is known that there are two types of $\gamma-$ray emitting active galactic nuclei (AGN): blazars and radiogalaxies. Their spectral energy distribution (SED) has typically two broad peaks: one, at low frequencies, is due to the radiation emitted by synchrotron processes; the second, at high frequencies, is thought to be due to inverse-Compton scattering (IC) of high-energy electrons off ambient seed photons. The underlying physical mechanism generating such a type of SED is supposed to be the same: a relativistic jet observed with different viewing angles, very small in the case of blazars and larger for radiogalaxies (see Fossati et al. 1998 and Donato et al. 2001 for blazars; see Ghisellini et al. 2005, Tavecchio \& Ghisellini 2008 for radiogalaxies). 

The SEDs of blazars seem to have different shapes depending on the emitted power and are organized in the so-called ``blazar sequence'' (Fossati et al. 1998). In particular, high-luminosity blazars have the peaks at low frequencies (``red'' blazars), while as the luminosities of the objects decrease, the peaks shift to higher frequencies, so that the lowest luminosity blazars are detected even at TeV energies (``blue'' blazars).

The blazar sequence can be interpreted in terms of changes of the seed photons for the  IC processes (Ghisellini et al. 1998). Blue blazars have no or weak emission lines (equivalent width $EW < 5$~\AA) and the IC seed photons are those from the synchtrotron radiation (e.g. Ghisellini et al. 1985, Band \& Grindlay 1985). Instead, red blazars have strong ($EW > 5$~\AA) emission lines and the seed photons are from the broad-line region (BLR) or accretion disk or even from the molecular torus (e.g. Dermer et al. 1992, Sikora et al. 1994, B{\l}a\.zejowski et al. 2000). It is worth noting that all the permitted emission lines - with weak or strong intensities - are broad, i.e. with $FWHM > 2000$~km/s (see, e.g., Wills \& Browne 1986, Wang et al. 2009).

This was, roughly speaking, the scenario before of the launch of \emph{Fermi}. 

\section{The discovery of gamma-rays from PMN~J0948+0022}
The surprise came with the detection by the Large Area Telescope (LAT, Atwood et al. 2009), onboard the \emph{Fermi} satellite, of a bright $\gamma-$ray source associated with the quasar PMN J0948+0022 (Abdo et al. 2009a,b,c). This quasar is known to be a radio-loud narrow-line Seyfert 1, with narrow permitted lines, bump of FeII and the flux ratio between [OIII] and $H\beta$ smaller than 3 (Zhou et al. 2003, Komossa et al. 2006, Yuan et al. 2008). Particularly, the FWHM of $H\beta$ is about $1500$~km/s (Zhou et al. 2003, Yuan et al. 2008), the narrowest permitted line ever detected in a $\gamma-$ray emitting AGN. Two observations at different epochs (28~February and 27~March~2000) were available at the \emph{Sloan Digital Sky Survey} (SDSS\footnote{\texttt{http://www.sdss.org/}}), indicating a change in the intensity (EW) of the $H\beta$ from $16$~\AA \, to $21$~\AA \, within about one month. 

This ``surprise'' was somehow expected, since radio observations of PMN J0948+0022 have shown a compact source, with flat spectrum and high brightness temperature, suggesting the presence of a relativistic jet (Doi et al. 2006, Yuan et al. 2008). The $\gamma-$ray detection by \emph{Fermi} confirmed this suggestion and allowed to build a complete SED: it was found that PMN J0948+0022 has the characteristics of flat-spectrum radio quasars (FSRQ), but with low power, relatively small mass ($1.5\times 10^{8}M_{\odot}$) and high accretion ($40$\% the Eddington value). More details can be found in Abdo et al. (2009c). We take the opportunity of this work to deal with more details on the classification. Indeed, this will be the first source of a new population of $\gamma-$ray emitting AGN and, therefore, it is necessary to address carefully all the known possible issues and doubts. 

\begin{figure}[!t]
\centering
\includegraphics[scale=0.5,clip,trim = 0 50 0 40]{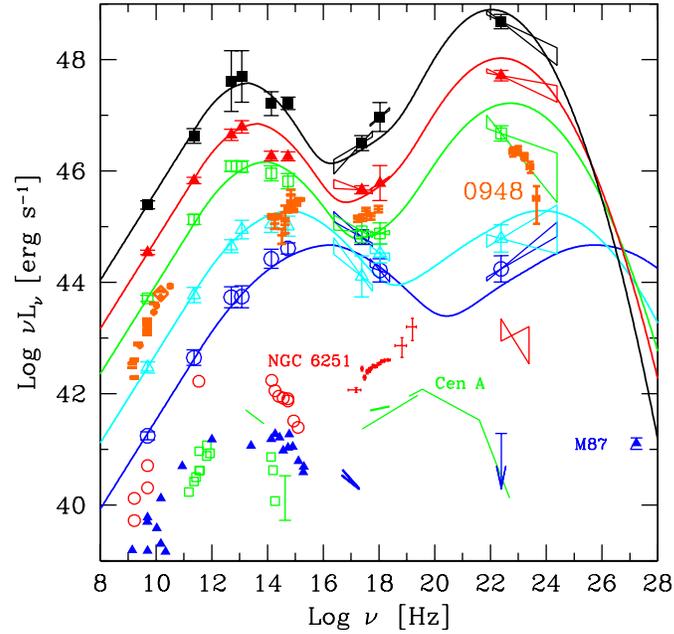}
\caption{Comparison of the SED of PMN J0948+0022 (from Abdo et al. 2009c) with the spectral sequence of blazars and with SED of some well-known powerful radiogalaxies. Adapted from Ghisellini et al. (2005).}
\label{fig:SED}
\end{figure}

\begin{figure}[!h]
\centering
\includegraphics[angle=270,scale=0.36]{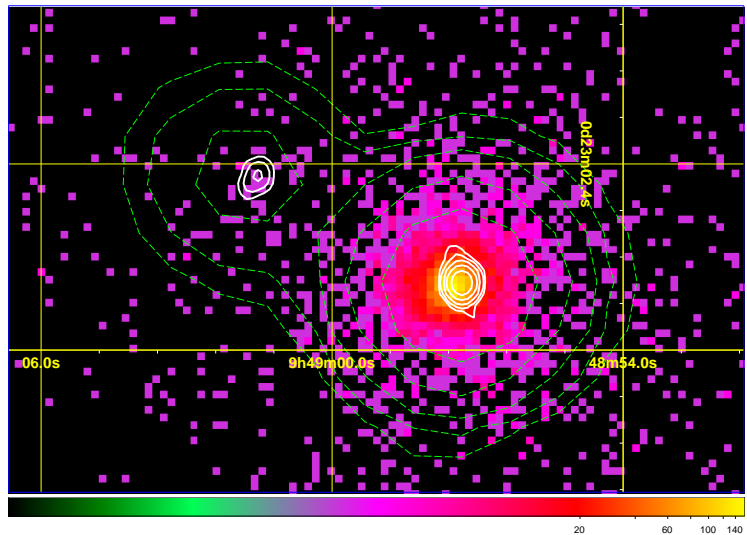}
\caption{X-ray image ($0.2-10$~keV) of PMN J0948+0022 obtained by integrating $12$ Swift/XRT observations performed between $2008$ and $2009$, for a total exposure of $54.4$~ks. Radio observations at $1.4$~GHz from NVSS (dashed yellow lines) and FIRST (continuous white lines) are superimposed. Epoch of coordinates is J2000. Color bar indicates X-ray counts.}
\label{fig:XRTFIRST}
\end{figure}

\section{Differences and similarities with blazars and radiogalaxies}
Fig.~\ref{fig:SED} shows the SED of PMN J0948+0022 compared with the blazar sequence (continuous lines of different colors) and a few of the most powerful radiogalaxies (Cen~A, M~87, NGC~6251). It is immediately evident that PMN J0948+0022 is in the region of blazars, with the observed emitted power well above the radiogalaxies region. This is an observational evidence, which does not need for any further explanation.

The study of the morphology of the source at radio frequencies shows a very compact source from $1.7$ to $15.4$~GHz (see Doi et al. 2006), except for the data at $1.4$~GHz from the NVSS (Fig.~\ref{fig:XRTFIRST}). In this case, there seems to be an extended structure, but it is likely to be an artifact of low angular resolution of NVSS ($FWHM=45''$, Condon et al. 1998). Indeed, images at higher resolution ($FWHM=5''$, Becker et al. 1995) from the FIRST survey ($1.4$~GHz) indicated the presence of two resolved sources, $1'.2$ distant each other, one of which is the core of PMN~J0948+0022 (with flux $107.5\pm 0.1$~mJy), while the second one is unknown and has a radio flux of $8.0\pm 0.1$~mJy. 

No optical data are available in any public catalog. \emph{Swift}/UVOT observations found no source at any filter with these upper limits ($3\sigma$, in units of $10^{-13}$~erg~cm$^{-2}$~s$^{-1}$): $V < 3.7$ , $B < 2.5$, $U < 1.5$, $UVW1 < 0.9$, $UVM2 < 0.9$, $UVW2 < 0.6$. No X-ray source was found by integrating all the available $12$ Swift/XRT observations (total exposure $54.4$~ks), with an upper limit ($3\sigma$) of $1.4\times 10^{-14}$~erg~cm$^{-2}$~s$^{-1}$ in the $0.2-10$~keV energy band. 

Search for detections at lower frequencies did not result any useful data (A. Capetti, private communication). No detection was found, both for PMN J0948+0022 and the unknown nearby source in the Very Large Array Low-frequency Sky Survey (VLSS) at $74$~MHz (Cohen et al. 2007), with an upper limit of $300$~mJy, which is not very constraining. 

It is therefore unlikely that there is any link between PMN J0948+0022 and the unknown source. We can also add that - in case we considered the unknown source as extended emission of the RL-NLS1 - the ``extended''/core flux ratio of 0.13 is in the range of lobe-dominated sources (cf Ghisellini et al. 1993), which in turn imply a strong X-ray emission from the unknown source (not observed) and very low Doppler factors ($\delta \approx 1$) at odds with $\gamma-$ray emission and the observed variability.

The most striking differences with blazars and radiogalaxies emerge in the optical spectrum. Again, since radiogalaxies have optical spectra similar to Seyferts or Low-Ionization Nuclear Emission-Line Regions (LINERs), it is tempting to try recovering this option. However, this is not the case.

LINER are easily discarded because of the high-power of the source (cf Fig.~\ref{fig:SED}) and the absence of other low-ionization lines, such as [OI] (for a review, see Ho 2008). With respect to the differences with Seyferts, NLS1 are indeed an AGN class separate from Seyfert 1 and 2 (for a review, see Pogge 2000). As known, it is possible to observe broad (permitted) and narrow (permitted and forbidden) emission lines in Seyfert 1, while only narrow lines can be observed in Seyfert 2, since a molecular torus on the line of sight hampers the viewing of the broad-line region (BLR) and allow only to see the narrow-line region (NLR). In the case of NLS1, we are observing lines from the BLR, which are narrower than usual, but are indeed coming from the BLR, not from the NLR (see also Rodr\'iguez-Ardila et al. 2000). There is no obscuration hampering the viewing of the BLR, as in Seyfert 2, otherwise this would result in [OIII]/H$\beta > 3$ (like in Seyfert 2) and the absence of FeII bump, which is observed in Seyfert 1, but not in Seyfert 2. 

The narrowness of the permitted lines is an indicator of physical conditions, which are really different from those in other Seyferts. Decarli et al. (2008) suggested that the observed spectrum of NLS1 is due to the fact that these sources have a disk-like BLR and we are observing it pole-on. Therefore, there is no component of the circular motion in the disk directed toward the observer, which can cause the Doppler broadening. On the other hand, Marconi et al. (2008) explain the narrowness of BLR lines as due to the radiation pressure of an accretion disk close to the Eddington limit, which pushes the BLR farther from the central spacetime singularity. 

\section{The morphology of the host galaxy}
Due to the high-redshift of PMN J0948+0022 ($z=0.585$), no direct information of the morphology of its host galaxy is available. However, since Seyferts are generally hosted by spiral galaxies and radio-loud AGN are in ellipticals, the possibility to find a relativistic jet in a spiral galaxy is surely charming. There are some studies on the NLS1 morphology suggesting a bulge-dominated structure (Zhou et al. 2006), even with some starburst contribution (Ant\'on et al. 2008 Sani et al. 2009). Work is in progress to study the specific case of PMN J0948+0022.

\section{Conclusions}
We can conclude that PMN J0948+0022 is a $\gamma-$ray emitting AGN really different from blazars and radiogalaxies and, therefore, \emph{it represents the first detection of a likely emerging new population of $\gamma-$ray AGN}. 

It is not clear what is the impact of these newly discovered sources in the unified scheme of AGN and, specifically, for radio-loud sources. For example, such a type of source was not predicted by the well-known unified scheme of radio-loud AGN proposed by Urry \& Padovani (1995). Another example is in the Fig.~7 of Boroson (2002), where this population would hardly find its place: it should be in the right part of the figure, in the radio-loud region, perhaps at the border with radio-quiet objects and close to the BAL QSO (Broad-Absorption Line Quasi-Stellar Objects) region, but if it will be confirmed a spiral host galaxy for RL-NLS1, this would be at odds with the whole diagram. 

So, we are not able yet to give an answer to these questions and issues. A multiwavelength campaign on PMN J0948+0022 was performed from 26 March to 5 July 2009 and will surely give some additional hints to understand the nature of RL-NLS1. More information will likely to come from the increasing numbers of this type of sources that will be detected at $\gamma-$rays by \emph{Fermi}/LAT.

\acknowledgements 
The \emph{Fermi} LAT Collaboration acknowledges support from a number of agencies and institutes for both development and the operation of the LAT as well as scientific data analysis. These include NASA and DOE in the United States, CEA/Irfu and IN2P3/CNRS in France, ASI and INFN in Italy, MEXT, KEK, and JAXA in Japan, and the K.~A.~Wallenberg Foundation, the Swedish Research Council and the National Space Board in Sweden. Additional support from INAF in Italy for science analysis during the operations phase is also gratefully acknowledged.

This research has made use of data obtained from HEASARC, provided by NASA/GSFC, and of the NASA/IPAC Extragalactic Database (NED) which is operated by the JPL, Caltech, under contract with the NASA.

EA acknowledges the use of the $100$-m telescope of the MPIfR (Max-Planck-Institut f\"ur Radioastronomie) at Effelsberg.

\end{document}